\documentstyle[aps,prl,psfig,multicol]{revtex}

\begin{document}

\date{UFIFT-HEP-97-9 $~~~~$ January 1998}

\vspace{0.5in}
\title{Expanding Bubbles in a Thermal Background}

\vspace{1.cm}

\author{ Richard M. Haas\footnote{email: rhaas@phys.ufl.edu} }

\vspace{1.0cm}

\address{ Department of Physics,
University of Florida,
Gainesville, FL 32611}

\maketitle

\vspace{1.cm}

\begin{abstract}
\baselineskip 16pt

Real scalar field models incorporating asymmetric 
double well potentials will decay
to the state of lowest energy. While the eventual nature of the system 
can be discerned, the determination of the dynamics of the bubble wall
provides many difficulties. In the present study we investigate numerically
the evolution of spherically symmetric expanding bubbles coupled to a
thermal bath in $3+1$ dimensions. A Markovian Langevin equation is employed
to describe the interaction between bubble and bath. We find the 
shape and velocity 
of the wall to be independent of temperature, yet extremely sensitive
to both asymmetry and viscosity.


\vspace{0.5cm}

\noindent PACS: 11.10.Lm, 05.70.Lm, 98.80.Cq

\end{abstract}

\begin{multicols}{2}
\narrowtext

\section{Introduction}

The behavior of fields during weak first order phase transitions has 
been a source of great interest.
Several studies have shown either a need for changes to the current 
theory of thermal phase transitions or the inclusion of
carefully calculated corollaries to the present models
\cite{phase_mix,oscillon,thermo_osc}.
Sub-critical bubbles and oscillon solutions may give 
rise to phase mixing in which these fluctuations in the smooth homogeneous
background can have significant dynamical effects. Thus in the limit of weak
first order phase transitions, the assumption of small amplitude fluctuations
may need revision. For stronger transitions, the fluctuations are expected 
to diminish in importance validating the use of standard homogeneous nucleation
theory. Yet, in the realm of the expanding bubble, precious few 
analytic and numerical studies have been performed to determine 
the effect of the thermal bath on the expansion of the bubbles
\cite{lower_dim,2d_sim_expand}.
With this in mind, we numerically examine expanding bubbles in $3+1$
dimensions using a Langevin equation 
in hopes of gaining either qualitative or quantitative results 
that describe their behavior.

Symmetry restoration at various stages in the Universe's life has
become an important and powerful concept in modern cosmology.
Several examples of the relevance of phase transitions are given by 
inflation \cite{inflation}, the electroweak phase transition 
\cite{ewpt}, and the quark hadron phase transition \cite{qhpt}.
In particular, the phase transition that occured when the 
universe cooled to a critical temperature of about $300$ GeV 
broke the symmetry of the weak and electromagnetic interactions.
The breakdown of SU(2) $\times$ U(1) which may have created the baryon 
asymmetry we see today is expected to be of first order
\cite{first_ord_break}. This satisfies the third of Sakharov's
conditions for baryogenesis, namely that no thermal equilibrium 
exist \cite{sakharov}.

At the electroweak scale, the rate of 
expansion of the universe is such that thermal equilibrium is
maintained \cite{kolb_turner}. To create an out of equilibrium process,
a bubble of true vacuum appears within the false vacuum state.
As the bubble expands, baryogenesis takes place in the neighborhood
of the boundary of the bubble, the bubble wall.
Thus, the detailed behavior of the bubble wall is of great 
importance in the electroweak phase transition.

The nucleation of bubbles in the context of field theory has by
now a long history.
Coleman has shown that at zero temperature,
the transition to the true vacuum takes place via the quantum
nucleation of bubbles \cite{coleman}. 
This process corresponds to a generalization of barrier penetration
in quantum mechanics.
The probability for bubble nucleation at zero temperature 
is found by determining the ``bounce'' configuration from
the Euclidean equation of motion subject to the boundary conditions of 
restricting the field to the metastable minimum at $t_{E} = - \infty$,
the global minimum at $t_{E} = 0$, and the metastable minimum at
$t_{E} = + \infty$.
The bounce solution is used to compute the Euclidean action, from
which bubble nucleation probability per unit volume is determined,
$\Gamma = A \exp{ ( - S_{E} ) }$.
The dominant contribution to the tunneling rate comes from the 
solution to the equation of motion with the least action.
These solutions are $O(4)$ symmetric, satisfying a $O(4)$
symmetric Euclidean equation of motion.
The nucleating bubbles were shown to quickly accelerate to the 
speed of light.

Linde expanded Coleman's ideas to account for finite-temperature 
effects in the nucleation of bubbles \cite{finite_linde}.
Temperature effects can be included by formal substitution of the
Euclidean time to include temperature, imposition of periodic boundary 
conditions on the field, and integration in the ``new'' Euclidean time
from 0 to 1 \cite{ramond:book}.
For high temperatures, the kinetic term in the action grows large. 
Yet, as solutions which minimize the action are sought,
the time independent configurations will dominate.
The $O(3)$ Euclidean equation can be solved to find the finite 
temperature bounce. 
The rate of thermal nucleation of critical bubbles per unit volume
was found to include an ``activation energy'' term representing the 
free energy barrier the system must overcome for the transition
to occur. 
The thermal fluctuations were shown to be a purely classical effect.
Thus the previous results of the bounce solutions also apply to the
finite-temperature case with different configurations for different 
temperatures.

Recent numerical studies have shown that the actual first order phase 
transition from false to true vacuum in the presence of a heat bath 
may be a great deal more complicated than indicated from the picture of 
a finite temperature bounce solution.
For weak first order phase transitions, phase mixing through
sub-critical bubbles may effectively restore the symmetry \cite{phase_mix}.
Long-lived sub-critical bubbles, oscillons, may act as 
nucleation sites for critical bubbles, thus speeding the 
transition rate \cite{oscillon,thermo_osc}.
Although the importance of these effects are still somewhat in
dispute \cite{dispute},
it is clear that significant deviations from the standard model of 
bubble nucleation may occur and that carefully considered numerical
calculations must be performed in order to determine the behavior
of the system. 

Previous studies of thermal nucleation of bubbles have dealt with
systems in 
lower dimensions. Work in 1+1 dimensions has primarily focused on the 
nucleation of kink-antikink pairs \cite{1d_sim}. The behavior that was
examined was the kink density and lifetime for a wide range of viscosities 
and temperatures.
The numerical results showed good agreement with the 
theoretical expectations.
In 2+1 dimensions, studies have focused on the validity of theoretical 
calculation of the nucleation barrier when compared to numerical 
results \cite{2d_sim} and the effect of fluctuations on the structure
and velocity of the expanding bubble wall \cite{2d_sim_expand}.

In this study, we will investigate the behavior of a $3+1$ dimensional
bubble expanding in a thermal background.
To insure uncontaminated results, the field is localized in the metastable
vacuum with the heat bath acting until a bubble is formed.
Although this requires longer computational times, 
it provides results which occur naturally from the equations of motion
without need for an artificial contrivance.
We hope to shed light on two vital characteristics of the 
transition from our study
as the asymmetry of the potential, the bath's viscosity, and the bath's
temperature are changed:
1. the shape of the bubble wall, and,
2. the speed at which the bubble wall travels.

The rest of this work is organized as follows.
In the next section, we briefly review the properties that give rise
to expanding bubbles and the bounce solutions found by
Coleman \cite{coleman}.
In Section III, we discuss the effects of coupling the system to 
a thermal bath by generalizing the equation of motion found in 
section II using a Langevin form.
In Section IV, we present the numerical results.
In Section V, we summarize our results, pointing to possible directions 
for future work.

\section{Expanding Bubbles}

The behavior of a real scalar field follows from the form of 
the action,
\begin{equation}
S = \int d^4x \left[ \frac{1}{2}  (\partial_{\mu} \phi)^{2} 
	- V(\phi) \right ]
\; .
\end{equation}
Through use of Hamilton's principle, the classical equation of 
motion for the field is found to be
\begin{equation}
	\frac{\partial^{2} \phi}{\partial t^{2}} - \nabla^2 \phi =
	- {\partial V(\phi)\over \partial \phi}
\; .
\label{eom:eq}
\end{equation}
To determine the energy of the field, we integrate in space over the
stress-energy tensor, 
\begin{equation}
E(\phi) = \int d^{3}x \, T^{00} =
\int d^{3}x \left[ 
	\frac{1}{2} \left( \dot{\phi} \right)^{2} +
	\frac{1}{2} (\nabla \phi)^{2} + V(\phi) 
	\right]
\label{energy:eq}
\end{equation}
where the dot refers to partial differentiation with respect to time.
Thus, by specifying the potential, we will obtain an explicit equation 
of motion for the system and a determination of the energy of the field.

In the early universe, and, particularly, in the case of the electroweak
phase transition, the approximate expression for the effective potential
in the high temperature limit to one loop order is \cite{ew_pot_theory}
\begin{equation}
V(\phi, T) = D (T^{2} - T_{0}^{2}) \phi^{2} - ET \phi^{3} +
\frac{\lambda_{T}}{4} \phi^{4}
\label{eq:ew_pot}
\end{equation}
where
\[
D = \frac{1}{8 v_{0}^{2}} \left(
	2 m_{W}^{2} + m_{Z}^{2} + 2m_{t}^{2}
\right)
\]
\[
E = \frac{1}{4 \pi v_{0}^{3}} \left(
	2 m_{W}^{3} + m_{Z}^{3}
\right) \sim 10^{-2}
\]
\[
T_{0}^{2} = \frac{1}{2D} \left(
	\mu^{2} - 4 B v_{0}^{2}
\right) 
\]
\[
\lambda_{T} = \lambda - \frac{3}{16 \pi^{2} v_{0}^{4}} \left(
	2m_{W}^{4} \ln \frac{m_{W}^{2}}{a_{B} T^{2}} +
	m_{Z}^{4} \ln \frac{m_{Z}^{2}}{a_{B}T^{2}} -
	4 m_{t}^{4} \ln \frac{m_{t}^{2}}{a_{F} T^{2}}
\right)
\]
and
$2 \mu^{2} = m_{H}^{2}$,
$\ln a_{B} = 2 \ln 4\pi - 2 \gamma \simeq 3.91$,
$\ln a_{F} = 2 \ln \pi - 2 \gamma \simeq 1.14$.
Analyzing the temperature dependence of equation \ref{eq:ew_pot},
we find that for large temperatures the potential exhibits a single
unique minimum at $\phi=0$.
As the system evolves and cools, a non-global second minimum forms at 
a temperature of 
\[
T_{1} = \frac{ T_{0} }{
		\sqrt{1 - 9 E^{2} / 8 \lambda_{T_{1}} D }}
\; .
\]
With further cooling, degenerate minima of the potential form at 
the critical temperature $T_{c}$ given by
\[
T_{c} = \frac{T_{0}}{
		\sqrt{ 1 - E^{2}/ \lambda_{T_{c}} D} }
\; .
\]
The minima are located at
\[
\phi=0 \mbox{\hspace{0.3in} and \hspace{0.3in}}
	\phi_{c} = \frac{2 E T_{c}}{\lambda_{T_{c}}}
\; .
\]
When $T<T_{c}$ a new global minimum is formed. At $T=T_{0}$, the 
potential's global minimum is located at
\[
\phi = \frac{3 E T_{0}}{\lambda_{T_{0}}}
\]

For $T < T_{c}$, first order phase transitions are possible
for the system described by $V(\phi,T)$.
The cubic term in the potential is responsible for barrier formation
and the asymmetry between stable and metastable states.
The strength of the asymmetry depends upon $ET$.
To simplify the expression for equation \ref{eq:ew_pot} while retaining 
the parameters of physical interest
and allow for
possible applicability to other similar first order scenarios, the 
potential is written as
\begin{equation}
V(\phi) = \frac{m^{2}}{2} \phi^{2} - \frac{\alpha m}{3} \phi^{3}
	+ \frac{\lambda}{4} \phi^{4}
\; .
\label{eq:potadwp}
\end{equation}
The variable $\alpha$ will be a measure of the
asymmetry between the two minima.

To cast the system into a more tractable form, two further
steps can be taken to simplify the equation of motion.
By imposing spherical symmetry, we effectively reduce this problem to 
one dimension.
This allows for quicker computational manipulation and a clearer 
understanding of what is driving the system's behavior.
By neglecting azimuthal and longitudinal components, 
this study will serve to provide bounds on the behavior of the
expanding bubble.
Introducing the dimensionless variables 
$\rho=rm$,
$\tau=tm$,
$\Phi={{\sqrt{\lambda}}\over m}\phi$, and
$\tilde{\alpha} = \frac{\alpha}{\sqrt{\lambda}}$ the nonlinear 
equation of motion becomes
\begin{equation}
	{{\partial^2\Phi}\over {\partial\tau^2}}-{{\partial^2\Phi}
	\over {\partial \rho^2}}
	-{2\over {\rho}}{{\partial \Phi}\over {\partial \rho}}  = 
	- \Phi + \tilde{\alpha} \Phi^{2} - \Phi^{3}
\; .
\label{equ_of_mot:eq}
\end{equation}

To solve the equation of motion for our system, we introduce three 
boundary conditions. The field initial starts at rest,
\[
	\dot{\Phi}(\rho, 0) = 0
\; .
\]
To avoid an undefined value at the origin, the spatial derivative
is fixed
\[
	\frac{d \Phi(0, \tau)}{d \rho} = 0
\; .
\]
Eventually the system will nucleate a bubble of true phase in the false one. 
Thus, at spatial infinity we expect the field to be in the false vacuum
state,
\[
	\Phi(\rho \rightarrow \infty, \tau) = \Phi_0
\; .
\]
%

\psfig{file=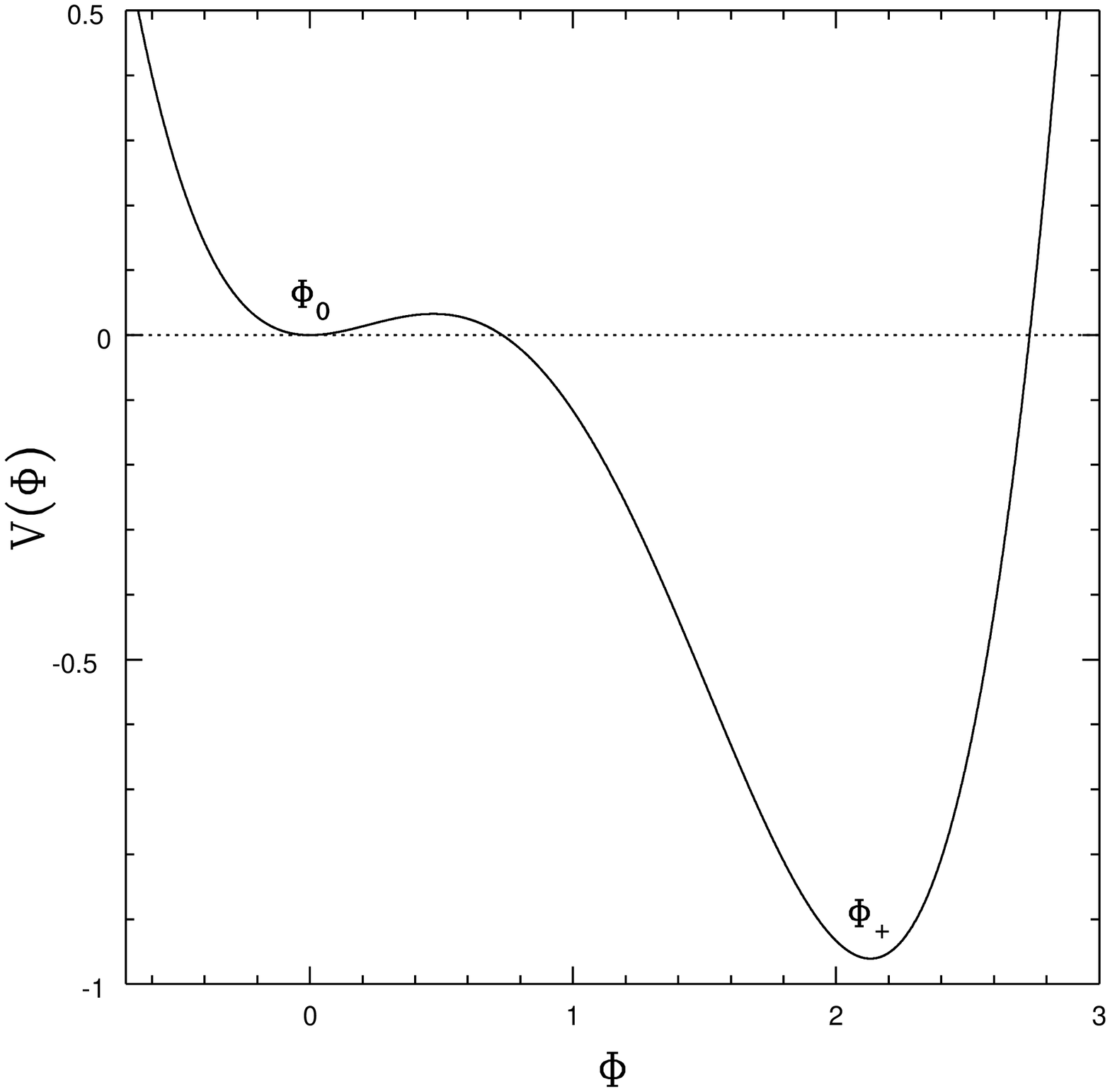,width=234pt}
\begin{figure}
	\caption{
	Asymmetric double well potential with $\tilde{\alpha} = 2.6$.
	}
\end{figure}

Using the dimensionless variables, the characteristics of the 
potential can be determined.
The minima of the potential are at
\[
\Phi=0 \mbox{\hspace{0.3in} and \hspace{0.3in}} 
\Phi_{+} = \frac{\tilde{\alpha}}{2}\left[ 1 +
\left( 1 - \frac{4}{\tilde{\alpha}^{2}} \right)^{\frac{1}{2}} \right]
\; .
\]
For $\tilde{\alpha} = \sqrt{\frac{9}{2}}$, the potential is 
degenerate. As shown in figure 1, when $\tilde{\alpha} > \sqrt{\frac{9}{2}}$,
$\Phi_{+}$ becomes the stable vacuum and the possibility for the
creation of expanding bubbles exists.
The classical turning point of the potential is found at
\begin{equation}
\Phi_{tp} = \frac{2}{3} \tilde{\alpha} \left[
	1 - \left( 1 - \frac{9}{2 \tilde{\alpha}^{2}} \right)^{\frac{1}{2}}
	\right]
\; .
\label{tp:eq}
\end{equation}

The zero temperature bounce \cite{coleman} and
the finite temperature bounce \cite{finite_linde}
solutions are needed in calculating the bubble nucleation rate.
To calculate the bounce, we write a Euclidean version of
equation \ref{eom:eq} through use of 
$t \rightarrow -i x^{0}$ and
$x^{2} = t^{2} - | \mbox{\bf x} |^{2} \rightarrow
	- (x^{0})^{2} - | \mbox{\bf x} |^{2} = - |x_{E}|^{2}$.
The boundary conditions restrict the field to be localized in one of
the minima of the asymmetric double well potential
at $x^{0} = \pm \infty$, hence the name ``bounce''.
The least action Euclidean solution will contribute most to the 
nucleation rate, having $O(4)$ symmetry.
Recalling that a connection between the path integral formulation 
of quantum field theory and the statistical mechanics partition function
can be found by the formal substitution of
$x^{0} \rightarrow \beta$ with $\beta = \frac{1}{k T}$,
defining the functions on a domain of length $\beta$ and periodically 
connected in time, the finite temperature action can be determined.
The functional integral for the action now contains a kinetic term of 
quadratic order in temperature. At high temperature, this term can
be neglected as we are looking for the solution which minimizes 
the action. In other words, the least action solution in finite 
temperature will be time-independent and have $O(3)$ symmetry.
The $O(3)$ spherically symmetric equation of motion is
\begin{equation}
\frac{d^{2} \Phi}{ d \rho^{2} } + \frac{2}{\rho} \frac{d \Phi}{d \rho} =
	\frac{\partial V({\Phi})}{\partial \Phi}
\; .
\label{eom_3symm:eq}
\end{equation}

For the asymmetric double well potential, no closed-form solution exists
to equation \ref{eom_3symm:eq}. 
However, approximate solutions can be determined if one restricts the
investigation to asymmetries that are small compared to the barrier 
height of the potential. For the dimensionless form of the potential
this amounts to $\tilde \alpha \rightarrow \sqrt{\frac{9}{2}}$. 
In this limit, the volume energy dominates the surface energy in equation
\ref{energy:eq} only when the radii of the bubbles are large. Additionally,
the transition between the vacua in $\Phi(\rho)$ occurs only in a 
small interval, 
thus giving rise to the name ``thin-wall'' approximation \cite{coleman}.
As the first order derivative term in equation \ref{eom_3symm:eq} is only 
appreciable in a small transition region and $\rho$ is large, we may 
approximate our equation of motion by
\begin{equation}
\frac{\partial^{2} \Phi}{\partial \rho^{2}} = 
	\Phi - \tilde{\alpha} \Phi^{2} + \Phi^{3}
\; .
\end{equation}
Solving this equation for $\Phi(\rho)$ we obtain the shape of the 
bubble in the thin-wall limit
\begin{equation}
\Phi = \frac{1}{\sqrt{2}} \left[
	1 - \tanh{ \left( \frac{\rho - R}{2} \right) }
	\right]
\; ,
\label{thin_shape:eq}
\end{equation}
with $R$ the bubble radius.
This solution will be of use later in our investigation when we 
compare the numerically generated bubble wall to the shape expected from
a slightly asymmetric double well potential.

\section{Boiling Bubbles}

To include the effects of a thermal background on the evolution
of the field, a generalized Langevin equation can be formulated
and solved numerically.
Employing an analogy between classical Brownian motion and 
quantum field theory, we find an equation of motion by 
adding a viscosity and a noise term to our original equation
of motion, equation \ref{eom:eq}. Assuming a Markovian heat bath, the viscosity
coefficient is related to the noise $\xi( {\bf x}, t)$ by the 
fluctuation-dissipation theorem
\begin{equation}
\langle \xi({\bf x},t)\xi( {\bf x'},t')\rangle =
	2 \gamma T \delta(t-t')\delta^3({\bf x} - {\bf x'})
\; .
\end{equation}
The coupling constants in this system are treated as 
free parameters.
Although more complicated nonlocal forms of the Langevin equation 
could be used \cite{nonloc},
the increased complexity in both the equation of motion and the dynamics 
do not warrant such an approach until the present situation is better
understood.
Indeed, the agreement between numerical studies and theoretical
predictions of nucleation in 1+1 dimensions lends support 
to the use of an additive noise and a Markovian bath \cite{1d_sim}.

The system is restricted to spherical symmetry.
As the thermal bath is provided with no symmetries, the proper 
evaluation of the system should contain all spatial dimensions.
Yet the advantages provided by evaluating the fully dimensional case
is questionable in light of the adversity this would cause.
The computer simulations of a non-symmetric system
would be expected to have significantly longer run times.
More importantly, the dominant
perturbations due to the thermal fluctuations are expected to be 
along the radial direction.
By allowing only variations in the radial direction, a clearly defined 
bubble wall velocity can be determined.
For these reasons, the values obtained here 
should be considered as an approximate bound on the velocities of
expanding (3+1)-dimensional bubbles.
Writing the fluctuation-dissipation theorem in terms of 
spherical symmetry, we obtain
\begin{equation}
\langle {\tilde \xi}(\rho,\tau){\tilde \xi}(\rho^{\prime},\tau')\rangle =
	\frac{1}{2 \pi \rho^{2}}
	{\tilde \gamma} \theta \delta(\tau-\tau')\delta(\rho - \rho^{\prime})
\; ,
\label{fluc_disp:eq}
\end{equation}
where $\theta = \frac{\lambda}{m} T$ is the dimensionless temperature.
The spherically symmetric dimensionless Langevin equation is written as
\begin{equation}
\frac{\partial^{2} \Phi}{\partial \tau^{2}} + 
	{\tilde \gamma} \frac{\partial \Phi}{\partial \tau} -
	\frac{\partial^{2} \Phi}{\partial \rho^{2}} -
	\frac{2}{\rho} \frac{\partial \Phi}{\partial \rho} = 
	- \Phi + {\tilde \alpha} \Phi^{2} - \Phi^3
	+ {\tilde \xi},
\label{langevin:eq}
\end{equation}
where ${\tilde \gamma} = \frac{\gamma}{m}$ 
and ${\tilde \xi} = \frac{\sqrt{\lambda}}{m^3} \xi$
are the dimensionless viscosity and noise respectively.

Equation \ref{langevin:eq} is numerically solved by a finite difference 
routine second order accurate in time and fourth order accurate in space.
The step sizes employed in the simulation were determined
rather crudely by specifying a numerically obtained bounce 
solution as the initial field configuration in the
$\tilde \gamma = 0$ case
and ensuring that this system behaved as expected for a 
sufficiently long time ($\tau > 6000$).
Satisfactory results were found in these situations if a 
spatial step of 
$\delta_{\rho} = 10^{-2}$ and a temporal step of 
$\delta_{\tau} = 2 \times 10^{-3}$ were used.

The validity of equations \ref{fluc_disp:eq} and \ref{langevin:eq}
can be tested by comparing numerical results and theoretical
predictions for the system subject to a quadratic potential.
Given sufficient time, the thermal evolution of the field in the 
potential will reach equilibrium. Since the system is described 
by a spatial one dimensional lattice of $N$ grid points, 
we can measure the energy per degree of freedom $E/N$. From basic
statistical mechanics, we know that once the system is in equilibrium
the equipartition theorem will give $E/N = \theta /2$.
The numerical results for this situation do show that the 
equipartition theorem holds for our system over a wide
range of parameters
[See Figure 2 of Ref. \cite{thermo_osc}].

While we have explicitly formulated a system whose dynamics are given
by the Langevin equation, many hydrodynamical studies have shed light
upon the phase transition \cite{hydro,hydro_instab,hydro_walvel}.
Whereas in the Langevin 
description the particles scattering off the bubble wall provide 
a damping proportional to the field's change in time, a hydrodynamic 
description treats the wall as a combustion front. 
The results of the hydrodynamic work have shown various characteristics
of expansion which are not manifest using the Langevin equation.
Instabilities have been found in the background fluid and the bubble wall 
which may dominate the dynamics \cite{hydro_instab}. 
The velocity of the bubble wall has also been found to be dependent 
upon the thermal conductivity of the background fluid and upon 
temperature inhomogeneities arising from the release of latent heat
\cite{hydro_walvel}. 
These considerations, while interesting and important, are not 
investigated in this work.
The dynamics of the bubble we obtain here come from the 
simplified description of the field subject to a potential
interacting with the heat bath by equation \ref{langevin:eq}.

\section{Numerical Results}

Equation \ref{langevin:eq} behaves in a manner consistent with 
expectations in the basic case of equilibrium thermodynamics. 
Computational grid step sizes have been found that reproduce
behavior present in specific aspects of nucleation. Our investigation
into the effects of a thermal bath on the shape and velocity of 
an expanding bubble wall may now begin.

For the field to be trapped in the metastable state and eventually 
decay by bubble nucleation to the global vacuum,
$\tilde \alpha > \sqrt{\frac{9}{2}}$. 
As shown in figure 2, the bubble is created from the background field
fluctuations. Eventually the fluctuations are large enough to form a 
critical bubble, a bubble that is large enough to complete the
phase transition and grow. 
To distinguish bubble formation from spurious background fluctuations
computationally, we examine the field searching for values greater
than the turning point. 
Once a segment of length $10$ in $\rho$ has been found whose members
satisfy this criterion, it is marked as an expanding bubble.

The shape of the bubble wall in figure 2 is reminiscent of the 
hyperbolic tangent thin-wall configurations.
Generalizing equation \ref{thin_shape:eq}, we obtain
\begin{equation}
\Phi(\rho, f) =
	\frac{\Phi_{+}}{2} \left[ 1 - \tanh \left( \rho f \right) \right]
\label{eq:bub_profile}
\end{equation}
where $\Phi_{+}$ is the global minimum and 
$f^{-1}$ specifies the thickness of the wall.
In figure 3a and 3b, equation \ref{eq:bub_profile} is used to fit 
expanding walls
generated in baths with 
$\tilde \alpha = 2.8$, $\tilde \gamma=1.0$, $T=3.0$ and
$\tilde \alpha = 3.1$, $\tilde \gamma=2.0$, $T=1.0$ respectively.
While the smoothness of the wall is altered when temperature is 
varied as asymmetry and viscosity remain constant, the bubble
wall's mean shape does not change.

\psfig{file=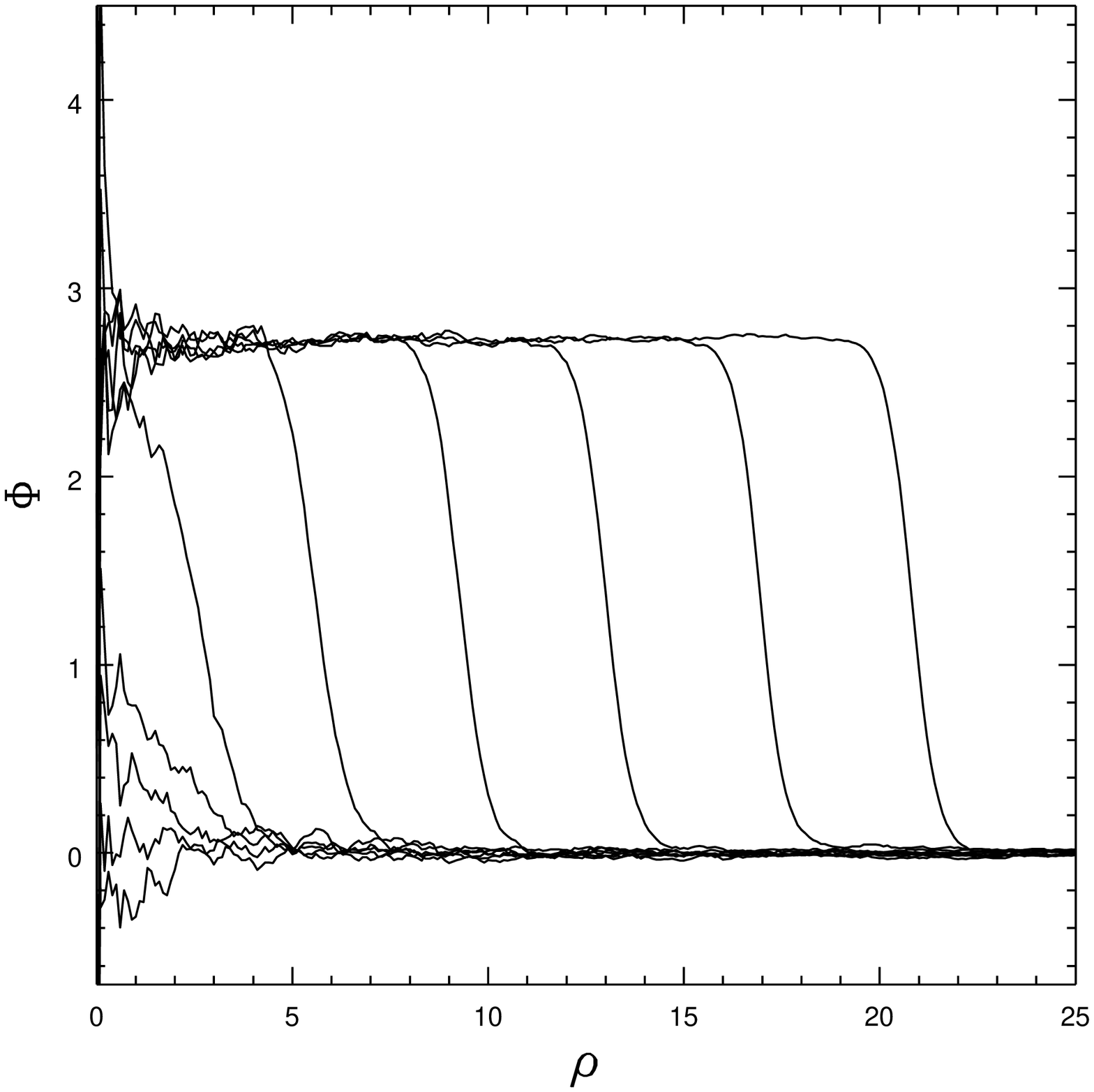,width=234pt}
\begin{figure}
	\caption{
		Snapshots of bubble evolution from the background field.
		The potential has an asymmetry $\tilde{\alpha} = 3.1$ and
		the heat bath is characterized by $\tilde{\gamma} = 1.0$ and
		$T=1.0$. The interval between snapshots is
		$\Delta \tau = 10^{-2}$.
	}
\end{figure}
\psfig{file=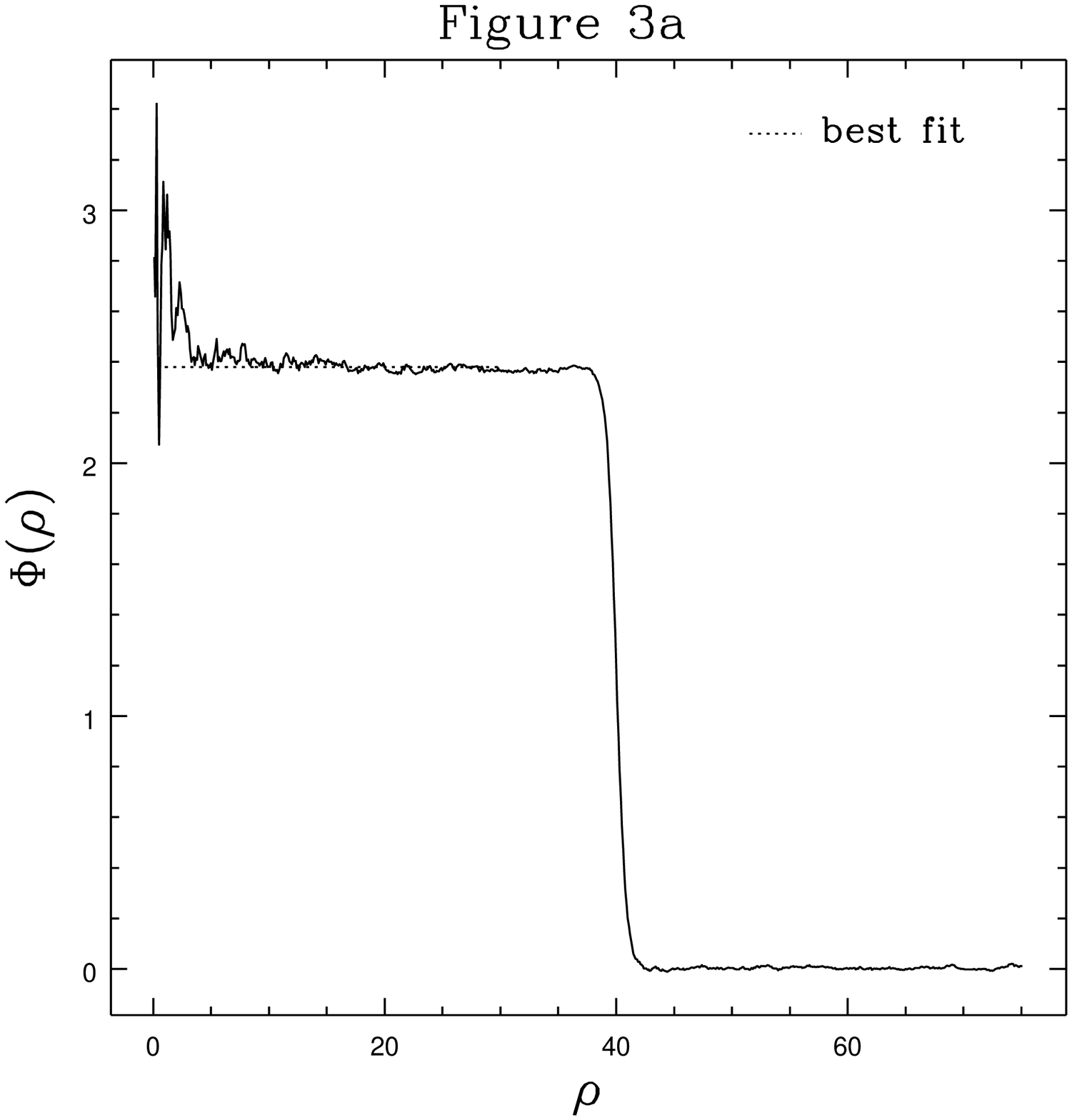,width=234pt}
\psfig{file=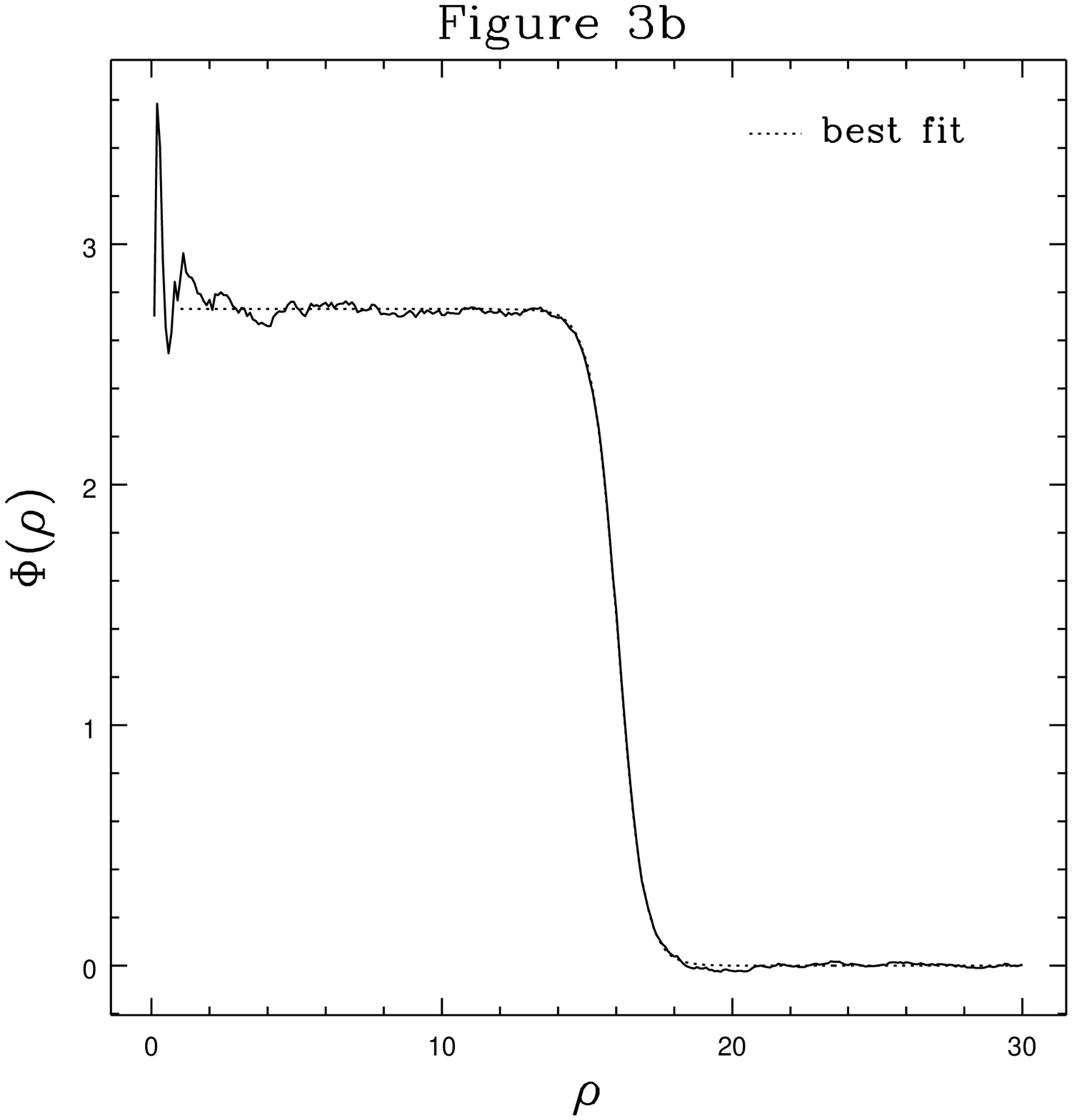,width=234pt}
\begin{figure}
	\caption{
		Snapshots of expanding bubbles with equation providing 
		best fit given by equation 14 for 
		(a) $\tilde{\alpha} = 2.8$, $\tilde{\gamma} = 1.0$, $T = 3.0$
		and (b) $\tilde{\alpha} = 3.1$, $\tilde{\gamma} = 2.0$, 
		$T = 1.0$.
	}
\end{figure}

The dependence of the wall thickness on the asymmetry of the potential 
and viscosity of the thermal bath can be found numerically by holding 
one quantity constant while varying the other.
For $\tilde \gamma = 1.0$ and $\tilde \alpha$ ranging between 
$2.4$ and $3.4$,
figure 4 shows that the data for $f$ can be described by a linear equation 
\begin{equation}
	f(\tilde \alpha) = A \tilde \alpha - B
\label{shap_asym:eq}
\end{equation}
with $A = 1.21$ and $B = 2.20$.
In figure 5, the numerical results of $f$ as asymmetry is held constant 
and viscosity varies is shown to be well fit by the equation
\begin{equation}
	f(\tilde \gamma) = A \tilde \gamma^{-\frac{3}{2}} + B
\label{shap_visc:eq}
\end{equation}
with $A = 0.512$ and $B = 0.969$.

\psfig{file=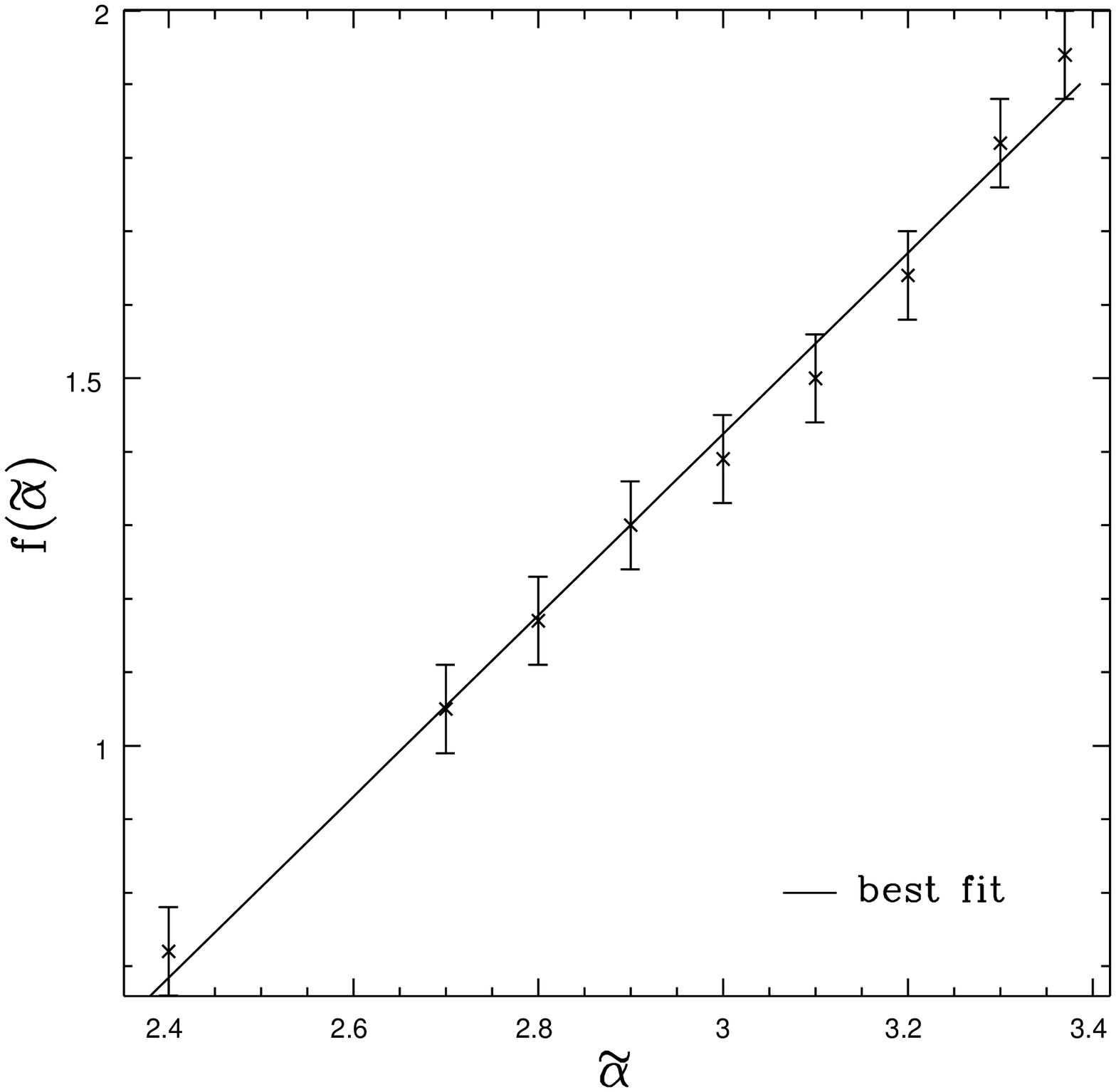,width=234pt}
\begin{figure}
	\caption{
		$f(\tilde \alpha)$ vs. asymmetry for constant
		viscosity $\tilde \gamma = 1.0$. The equation providing
		best fit is $A \, \tilde \alpha - B$ where
		$A = 1.21$ and $B = 2.20$.
	}
\end{figure}
\psfig{file=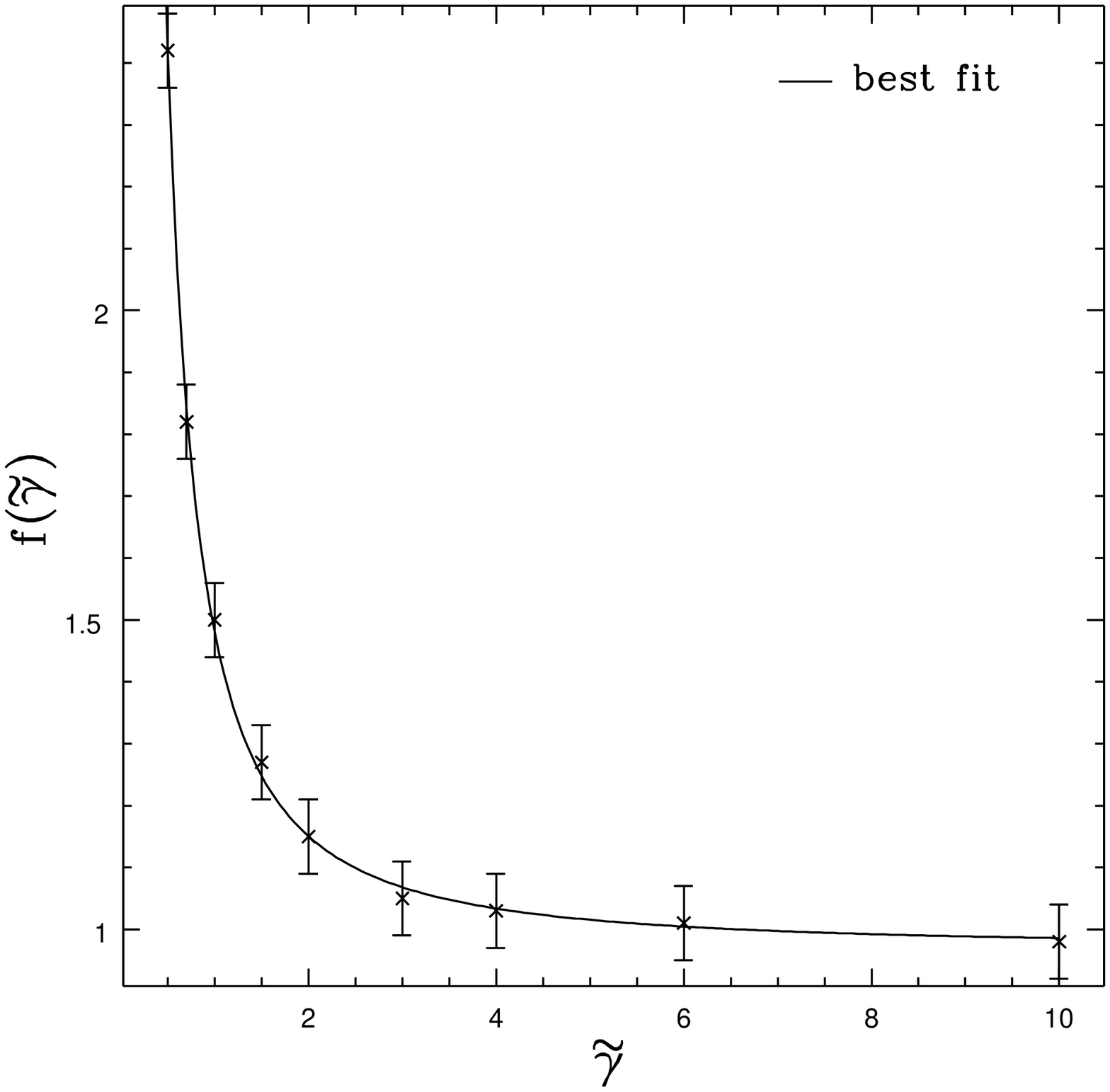,width=234pt}
\begin{figure}
	\caption{
		$f(\tilde \gamma)$ vs. viscosity for constant
		asymmetry $\tilde \alpha=3.1$.
		The equation providing best fit is 
		$A \, \tilde \gamma^{-\frac{3}{2}} + B$ where
		$A = 0.512$ and $B = 0.969$.
	}
\end{figure}

To ascertain the kinematics of the bubble wall,
a computational technique for determining the size of the bubble
must be developed.
As the turning point interpolates between the two vacua, 
the size of the bubble can be defined as that radial distance which
corresponds to the field configuration value being equal to the turning
point.
Recording the bubble size at a particular time allows one to find
the wall velocity through the equation 
$\upsilon_{bw} = \frac{\Delta \rho}{\Delta \tau}$
where $\Delta \rho$ is the difference between the bubble size at successive
times and
$\Delta \tau$ is the difference between the corresponding times.
In order to limit the already intensive use of computational resources 
employed in this study,
the simulations were terminated once the bubble reached a size of $80$.
Several trials with longer cutoff sizes were performed and found to yield
no additional benefit for the parameters probed.

The development of the bubble wall velocity through time 
is shown in figure 6. The graph clearly indicates a region in
which the bubble wall is accelerating, eventually reaching 
a terminal velocity.
Since the region of acceleration is only of a limited duration, we 
will ignore this, deferring a detailed analysis of the 
feature to later studies. 
Additionally, the restriction imposed of identifying an expanding
bubble by
a region of true vacuum phase of size 10 (in $\rho$) and greater
in most cases obscures the accelerating wall behavior. 
Thus,
although the data obtained exhibits an acceleration in the
start of a bubble's life, our computational methods restrict us 
to a purely qualitative analysis of this phenomena.

\psfig{file=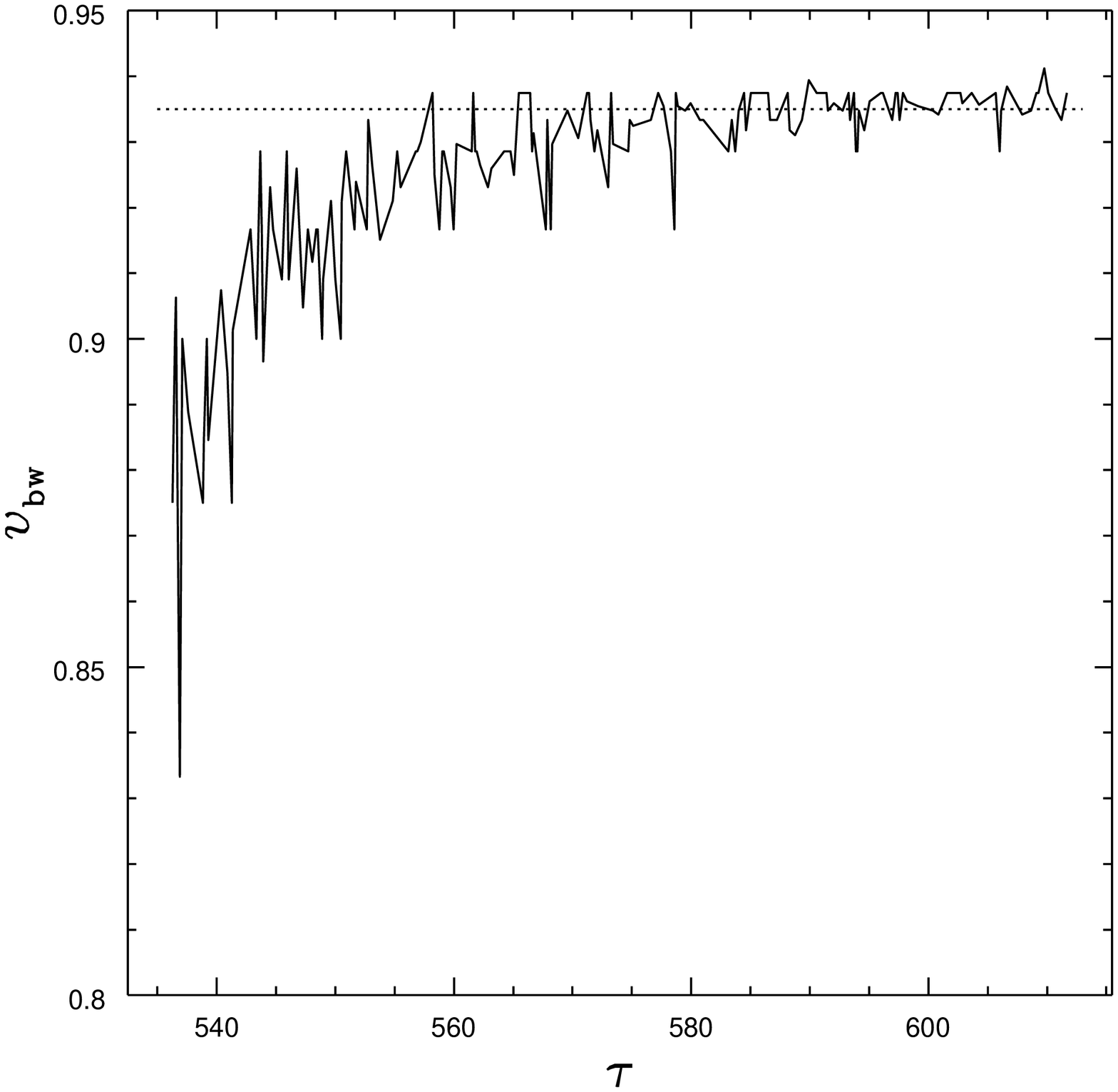,width=234pt}
\begin{figure}
	\caption{
		Bubble wall velocity vs. time for 
		$\tilde \alpha=3.1$, $\tilde \gamma=0.5$, and $T=0.5$.
		The dotted line indicates the constant 
		bubble wall velocity $0.935$.
	}
\end{figure}

The value for the velocity of the bubble wall is determined 
numerically through use of a ``windowing'' least squares method. 
The data over which a least squares fit for a linear equation was
performed was adjustable so as to find those portions which 
had a slope $| \frac{d \upsilon_{bw}}{d\rho} | \leq 10^{-5}$.
Once this criterion had been satisfied, the intercept was interpreted
as the terminal expansion velocity of the bubble.
For added certainty, a second method of smoothing the 
velocity as a function of time data 
using a Savitzky-Golay type filtering scheme was employed concurrently
\cite{NUM-REC}.
By examining the graph, the velocity could be visually discerned.
Both methods produced values in agreement with each other.

As in the case of the wall shape, the velocity of the wall was found 
to be invariant to the temperatures of the bath probed here 
($T=10^{-4}$ to $T=10$). 
To insure that this is not an artifact of some statistical dependence
in the noise term,
all computational simulations were given different random number seeds 
and a reliable random number generator used \cite{NUM-REC}. 
The generator was also checked for any correlations and none were found.
The parameters affecting the velocity were the bath viscosity and the 
potential asymmetry.

The bubble wall terminal velocity's dependence on asymmetry of the 
potential 
for constant viscosity $\tilde \gamma = 1.0$ is shown in figure 7.
The best fit equation to the numerical data is provided by
\begin{equation}
	\upsilon_{bw}(\tilde \alpha) = 
		\frac{1}{1 + A \tilde \alpha + B \tilde \alpha^{2}} + 1
\label{vel_asym:eq}
\end{equation}
with $A = 1.61$ and $B = -1.16$.

\psfig{file=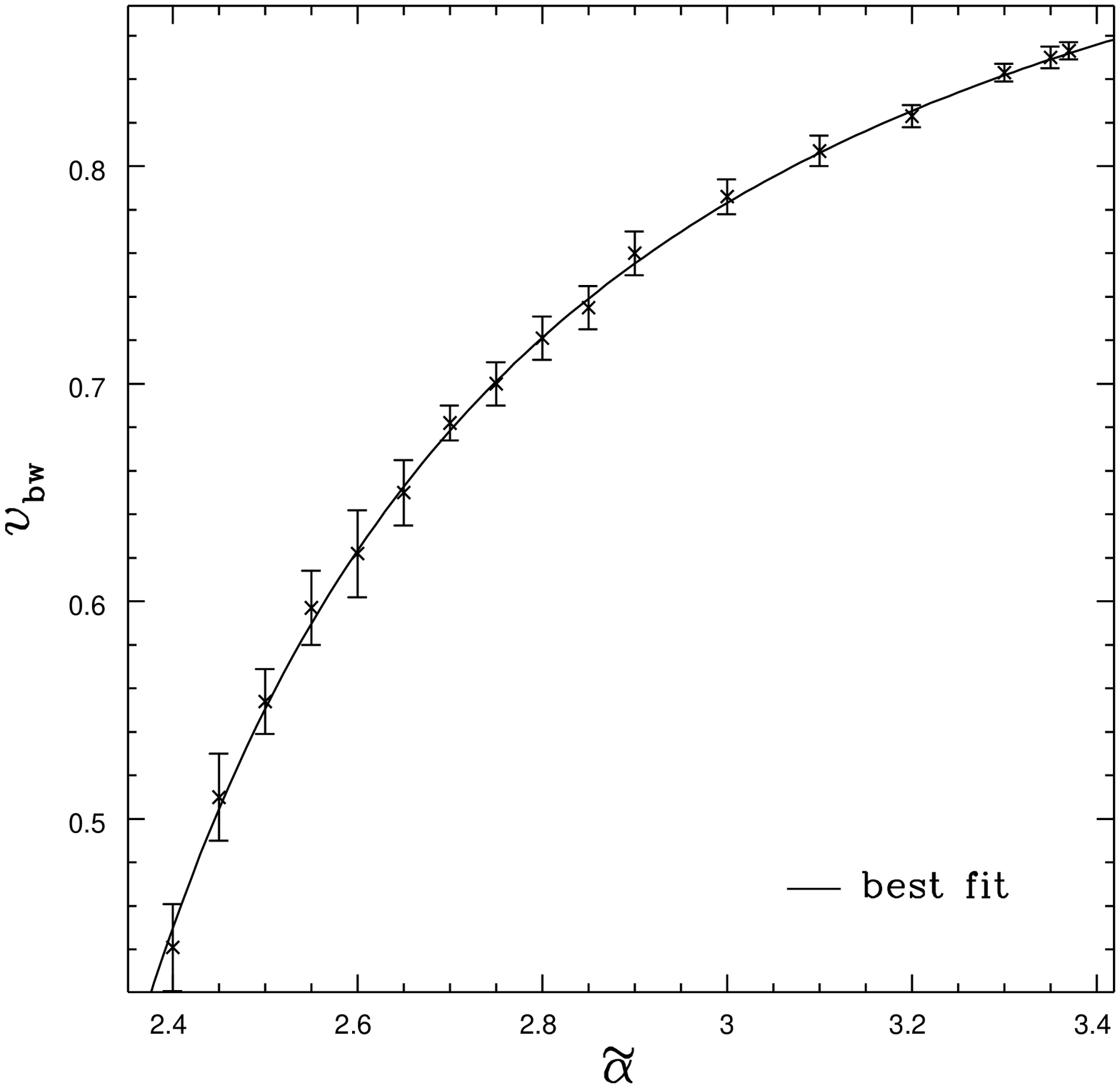,width=234pt}
\begin{figure}
	\caption{
		Bubble wall velocity vs. asymmetry for constant
		viscosity ($\tilde \gamma = 1.0$) and temperature
		($T = 1.0$). The equation providing best fit is
		$\frac{1}{ 1 + A \tilde \alpha + B \tilde \alpha^{2} } + 1$
		where $A=1.61$ and $B=-1.16$.
	}
\end{figure}

In figure 8, the asymmetry was fixed to $\tilde \alpha = 3.1$ while the 
viscosity was allowed to vary.
The velocity's dependence is well approximated by the expression
\begin{equation}
	\upsilon_{bw}(\tilde \gamma) = 
		\frac{A}{1+B \tilde \gamma^{C}} + D
\label{vel_visc:eq}
\end{equation}
with $A = 0.949$, $B = 0.275$, $C = 1.62$, and $D = 0.0584$.

\psfig{file=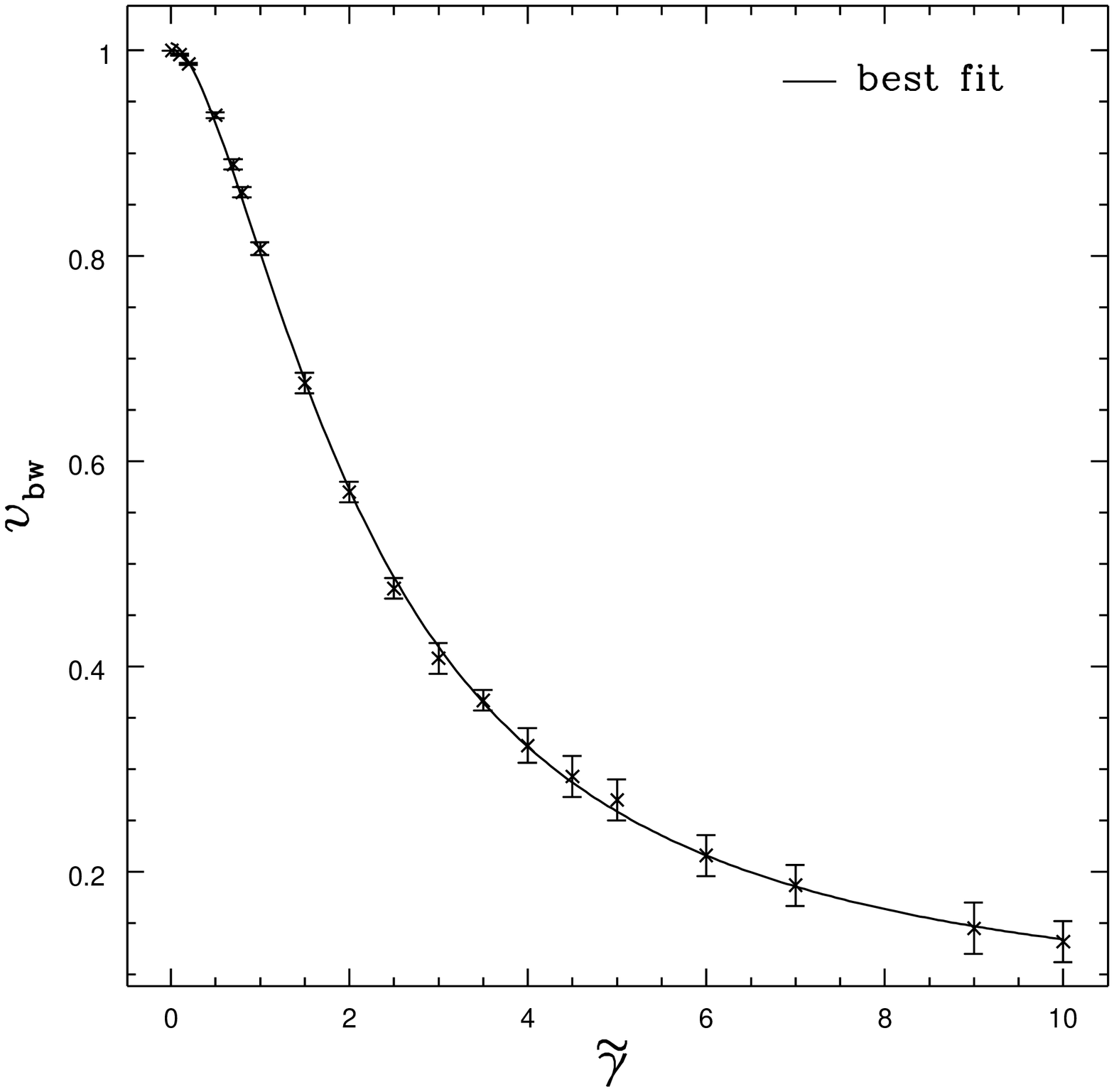,width=234pt}
\begin{figure}
	\caption{
		Bubble wall velocity vs. viscosity
		for constant asymmetry ($\tilde \alpha=3.1$)
		and temperature ($T=1.0$). The equation providing
		best fit is 
		$\frac{A}{1+B \tilde \gamma^{C}} + D$ where
		$A=0.949$, $B=0.275$, $C=1.62$, and $D=0.0584$.
	}                                                                    
\end{figure}

\section{Conclusions}

We investigated the shape and velocity of expanding bubble walls
in the presence of a thermal bath.
The dynamics of the system were modeled by a generalized Langevin
equation with a Markovian bath and an asymmetric double well potential.
The bubbles were generated naturally from the background thermal
field without need for specifying an initial configuration.
In the absence of the thermal bath, expanding bubbles eventually reach a 
terminal velocity of the speed of light and, in the thin wall limit, will 
have a profile corresponding to a hyperbolic tangent shape.

The presence of a thermal background strongly affects the shape and
velocity of the wall. 
While the relevant aspects of the system were invariant under 
changing temperatures, asymmetry and viscosity altered the dynamics of 
the expanding bubbles.
The shape of the wall is well approximated by the thin wall shape
with a thickness parameter that is dependent on both 
viscosity and asymmetry.
The inverse thickness of the bubble for changing asymmetries is
given by equation \ref{shap_asym:eq}.
In the presence of a heat bath, as the asymmetry increases, the
thickness of the bubble wall decreases.
For changing viscosity, the shape is given by equation 
\ref{shap_visc:eq}. The thickness of the bubble wall is found to increase
as the viscosity increases.
Unlike the expected results of increasing wall thickness as asymmetry grows 
for no thermal background \cite{kolb_turner},
the results here indicate that a relationship exists between viscosity which 
thickens bubbles and increasing asymmetry which tends to entice the system 
back to the expected results.
For velocity, the dependence on asymmetry is approximated by 
equation \ref{vel_asym:eq}. As the asymmetry of the potential increases
in the presence of a heat bath, the velocity will approach the expected
value of unity, the speed of light.
Yet, as is shown by equation \ref{vel_visc:eq}, the viscosity slows the
bubble wall.
Here the equation behaves as expected in the limit 
$\tilde \gamma \rightarrow 0$ giving light speed velocities for the 
expanding bubble wall.
If equation 18 is carried to large viscosity, the unsettling behaviour of
velocities for all viscosities exists. In fact, letting
$\tilde \gamma \rightarrow \infty$ one finds a suggestive lower bound
on expansion speeds of $0.0584$. This would indicate that viscosity
alone can never prevent the bubble from growing.
Yet, taking this result too seriously may lead to error due to the
lack of numerical trials of sufficiently large viscosities performed
and the methodology employed. It must be borne in mind that our
investigation has focused on only growing bubbles which evolve
naturally from the false vacuum. 
For large viscosities, the possibility of finding well-defined,
expanding bubbles becomes questionable. This possibility in turn may
lead to a discontinuity in an extended investigation of velocity
vs. time, eventually leading to the expected result of zero velocity.

Our investigation focused on spherically symmetric solutions to a 
Langevin equation with a Markovian bath.
Yet, there remain several questions whose answers would provide 
much new information on the subject of first order phase transitions.
To model truly real-world phenomena, ideally one would employ a fully
dimensional equation of motion. The prospects of doing this present serious 
challenges not only in the amount of present-day computational resources 
needed but also in such concepts as clearly defining a bubble wall velocity.
An interesting extension of this study appears to be in the
determination of the acceleration of the walls from bubble creation 
to the time of terminal velocity seen in the first few moments of figure 6.
More fundamental to the investigation of interacting field theories, 
it is possible that the thermal background will furnish
more complicated properties such as non-additive couplings and/or
nonlocal correlations in time and space \cite{nonloc}.
The consequences of these nonlocal effects are still questionable, but it 
is clear that their inclusion into the dynamics of the system by coupling to
more general thermal baths may have very 
significant effects on the shape and speed of the bubble walls. 
Of more fundamental theoretical importance is the possible connection
between the electroweak phase transition and the evolution of the
system according to the Langevin equation. While the methods employed
here are strongly suggestive of the electroweak phase transition, the
work of proving the connection between the two is left as an open question.

\acknowledgments

I would like to thank James Fry and Marcelo Gleiser for many useful 
discussions.
Additionally, I would like to thank John Yelton and the 
high energy experimental group at the University of Florida 
for access to many computer resources.
This work was also supported in part by the University of Florida
and the IBM Corporation through their Research Computing Initiative at 
the Northeast Regional Data Center.


\end{multicols}

\end{document}